# Sapphire Photonic Crystal Fiber Sensor

Mohan Wang, Tongyu Liu, Zipei Song, Richard Reeves, Frank P. Payne, Igor N. Dyson, Kaihui Zhang, Tao Wang, Jian Zhang, Zhitai Jia, Patrick S. Salter, Martin J. Booth, *Fellow, Optica*, and Julian A. J. Fells, *Member, IEEE*

***Abstract*—Sapphire optical fiber shows great promise for remote sensing in extreme environments approaching 2000 °C, by using laser-processing to form a single-mode waveguide within it. However, for practical application, longer devices with high manufacturability and reliability are required. We report the design, modeling, fabrication, and optimization of an index-guiding sapphire photonic crystal fiber Bragg grating temperature sensor. The device is fabricated using femtosecond laser direct writing to inscribe both the photonic crystal waveguide and the Bragg grating. A spatial light modulator was used to compensate for the mismatch between the immersion objective and the high-index oil used. This improved the aspect ratio and suppressed cracking during fabrication, for higher reliability. The design results in a 6-fold reduction in fabrication time over an equivalent depressed cladding waveguide, significantly reducing the cost of manufacture. Devices up to 7 cm long were fabricated and spliced to standard single-mode fiber. The propagation loss was estimated to be 0.7 dB/cm and the Bragg gratings had a bandwidth of approximately 0.12 nm. Devices were tested in a furnace showing a temperature sensitivity of between 19.0–32.3 pm/°C over a range 25-1200 °C. These longer devices have the potential to enable practical high precision extreme temperature monitoring in many applications, with lower manufacturing cost and higher reliability.**

## I. INTRODUCTION

OPTICAL fiber sensors are widely deployed for remote monitoring of infrastructure for improved safety, efficiency and reliability. However, emerging applications in energy, aerospace and industrial processing require sensing in more extreme environments than silica fiber can tolerate. Sapphire optical fiber is attractive as it can withstand ultra-high temperatures up to around 2000 °C, high radiation and corrosive chemicals, as well as having a wide spectral transmission window. However, a longstanding problem is that being coreless, it is multimoded [1]. This is particularly problematic for fiber Bragg grating (FBG) sensors as each mode has its own Bragg wavelength, resulting in a broad reflection spectrum, restricting resolution and multiplexing capability. The modal distribution is also sensitive to bending and environmental disturbances, further limiting sensing precision.

There has been considerable recent work to improve the performance of sensors in multimode sapphire fiber. Techniques used to excite a greater proportion of the modes include adding multimode fiber [2], [3] or using a mode scrambler [4], [5]. The number of modes launched into the sapphire fiber has been restricted by using a tapered silica fiber [6], or by fusion splicing to form a double conical structure [7]. Modes reflected from the FBG have been subsequently stripped away using offset coupling [8] or with a tapered waveguide in coreless silica fiber [9]. The number of modes the sapphire fiber supports has been restricted by using chemical etching to reduce its diameter [10] or by bending [11]. However, these approaches do not give perfect suppression of other modes and mechanical disturbance is likely to change the modes excited. There have been attempts to design the grating structure in such a way that it reflects a reduced number of modes. Such designs include a helical grating [12], a θ−shaped grating [13] and a decreasing ring grating [14]. However, these structures still require light to be within the modes that they reflect.

An alternative approach is to engineer the sapphire fiber such that it has a core and cladding. A lower refractive index cladding has been created within sapphire fiber by radiation exposure [15], while a hybrid growth technique has been used to create a 100 μm titanium doped core [16]. A photonic crystal fiber was formed from seven 70 μm fibers [17]. However, none of these fibers were single-mode. A single-mode sapphire fiber has been formed by using femtosecond laser direct writing to inscribe a depressed cladding structure and a Bragg grating within it [18]. These devices have shown very high repeatability at temperatures exceeding 1000 °C [19]. However, the high propagation loss of 1.5 dB/cm limited the device length to 4 cm. Furthermore, the design required 200-250 tracks, resulting in a very long fabrication time of 1-2 hours for a 4 cm device [19]. There were also cracks in the sapphire visible, likely due to the high proportion of the sapphire which was processed.

There is a need for longer devices in order to access ultrahigh temperature environments, whilst keeping the silica lead-out fiber within its operational temperature range. It is also important to reduce the fabrication time of the devices as this has a direct impact on the cost of manufacture. In this paper, we address these issues by developing a sapphire

This work was supported by the U.K. Engineering and Physical Sciences Research Council under Grant EP/T00326X/1. *(Corresponding author: Julian A.J. Fells).* Mohan Wang and Tongyu Liu are co-first authors.

Mohan Wang, Tongyu Liu, Zipei Song, Richard Reeves, Frank P. Payne, Igor N. Dyson, Patrick S. Salter, Martin J. Booth, and Julian A. J. Fells are with the Department of Engineering Science, University of Oxford, Parks Road, Oxford, OX1 3PJ, UK (e-mail: julian.fells@eng.ox.ac.uk).

Tao Wang, Kaihui Zhang, Jian Zhang, Zhitai Jia are with the State Key Laboratory of Crystal Materials, Shandong University, 27 South Shanda Road, Jinan, Shandong, 250100, China (email: z.jia@sdu.edu.cn).

Tao Wang is also with the Jiangsu Jingying Optoelectronics Technology Co. LTD, 10 East Zhujiang Road, Xuzhou, Jiangsu, 221116, China (email: t.wang@sdu.edu.cn).

Supplemental materials are provided.

Color versions of one or more of the figures in this article are available online at http://ieeexplore.ieee.org

photonic crystal fiber (PCF) design to lower the loss and extend the length, while also reducing the complexity and fabrication time.

## II. Design and Simulation

This work is an experimental realization of a concept for laser-fabricating a lattice waveguide within a single-crystal sapphire fiber to form a PCF [20]. Waveguides with sub-wavelength lattices have been demonstrated in YAG and sapphire using femtosecond laser–assisted chemical etching, but the high propagation loss limited lengths to around 1 mm [21]. Alternatively, photonic crystal geometries without a photonic bandgap, have been demonstrated in Nd:YAG [22], Nd:YAP [23], $LiNbO_3$ [24] and Ti:sapphire [25]. More recently, a 4 cm single-mode photonic crystal waveguide incorporating Bragg gratings was demonstrated in bulk sapphire [26].

Our design has an array of laser-modified tracks along the fiber, which are periodic in the cross-section. These tracks have a lower refractive index than the surrounding material. A core is formed by omitting tracks in the center of the lattice. Although there is no bandgap, the structure guides light by virtue of the lattice region acting as a cladding which has a lower average refractive index than the core. Such structures have been termed index-guiding photonic crystal waveguides and fibers [26], [27], [28], [29].

To implement the photonic-crystal structure within sapphire fiber, the feature size must be carefully controlled. Important parameters are: 1) the number of missing tracks in the lattice forming the core; 2) the lattice constant (axial separation between tracks); 3) the number of concentric layers of tracks surrounding the core; and 4) the refractive index change of the modification cross-section (magnitude, area and shape). Since there is a finite micro-structured cladding, the guided mode is intrinsically leaky: Even in the absence of material absorption, part of the field radiates into the cladding and escapes. We refer to this radiation-induced attenuation as the confinement loss [30], [31], [32]. From the propagation equation for the waveguide eigenmodes, we obtain:

$$E(x,y,z,t) = E_0(x,y,t)\mathrm{e}^{\mathrm{i}\beta z} \quad (1)$$

where $\beta = k_0 n_{eff}$ and $k_0 = 2\pi/\lambda_0$. The loss, $L$, can be written as:

$$L(dB) = \frac{20k_0\,\mathrm{Im}(n_{eff})z}{\ln 10} \quad (2)$$

From (2), we need only compute the imaginary part of the effective refractive index to obtain the confinement loss of the corresponding structure.

To identify an optimium design, we used COMSOL to simulate PCFs with different parameters and determine the effective refractive index, selecting a perfectly matched layer (PML) as the boundary layer to absorb all leaked electromagnetic waves without producing reflections at the boundary. In each simulation, the PML thickness and the mesh resolution were gradually increased to ensure convergence. We explored a range of lattice constants between 7 μm and 14 μm, and 2 to 6 concentric layers around the core. The simulations were performed for the 1550 nm wavelength. The bulk sapphire refractive index was taken as 1.746. Laser-induced refractive index changes between -0.003 and -0.005 were used, based on estimates from experimental work at different pulse energies. There is approximately a 50% reduction of the refractive index change after anneal [33], so it is important to use post-anneal figures for the design. The modification cross-sectional area and ellipticity were also taken into account, based on experimental observation.

We initially investigated a core formed by a single missing track, but this was found to be lossy. We also found that when a Bragg grating is inscribed within such a waveguide, the modification tracks of the grating are highly prone to physical overlap and interfere with the adjacent inner cladding modification lines. This spatial "crowding" perturbs the already weak waveguide boundary, leading to unacceptably high propagation losses. We therefore did not progress the configuration with one missing track, but the results are included in the Supplementary Information.

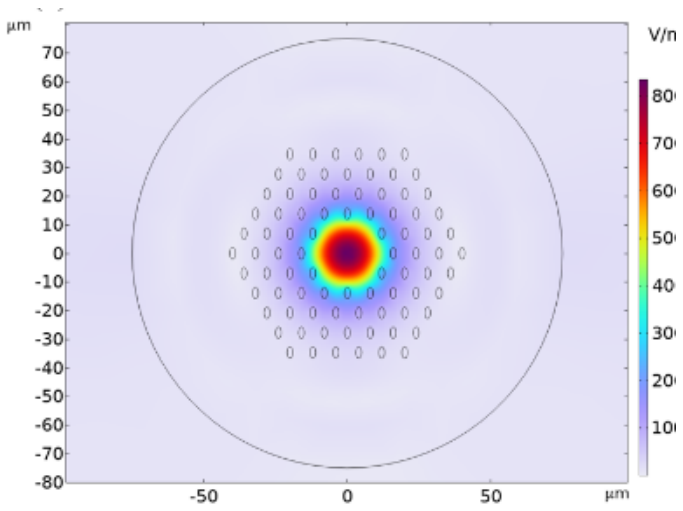

**Fig. 1** Example simulation mode profile of an index-guided sapphire PCF. The lattice constant is 8 μm, the core has 7 missing tracks, and there are 4 concentric layers of tracks around the core.

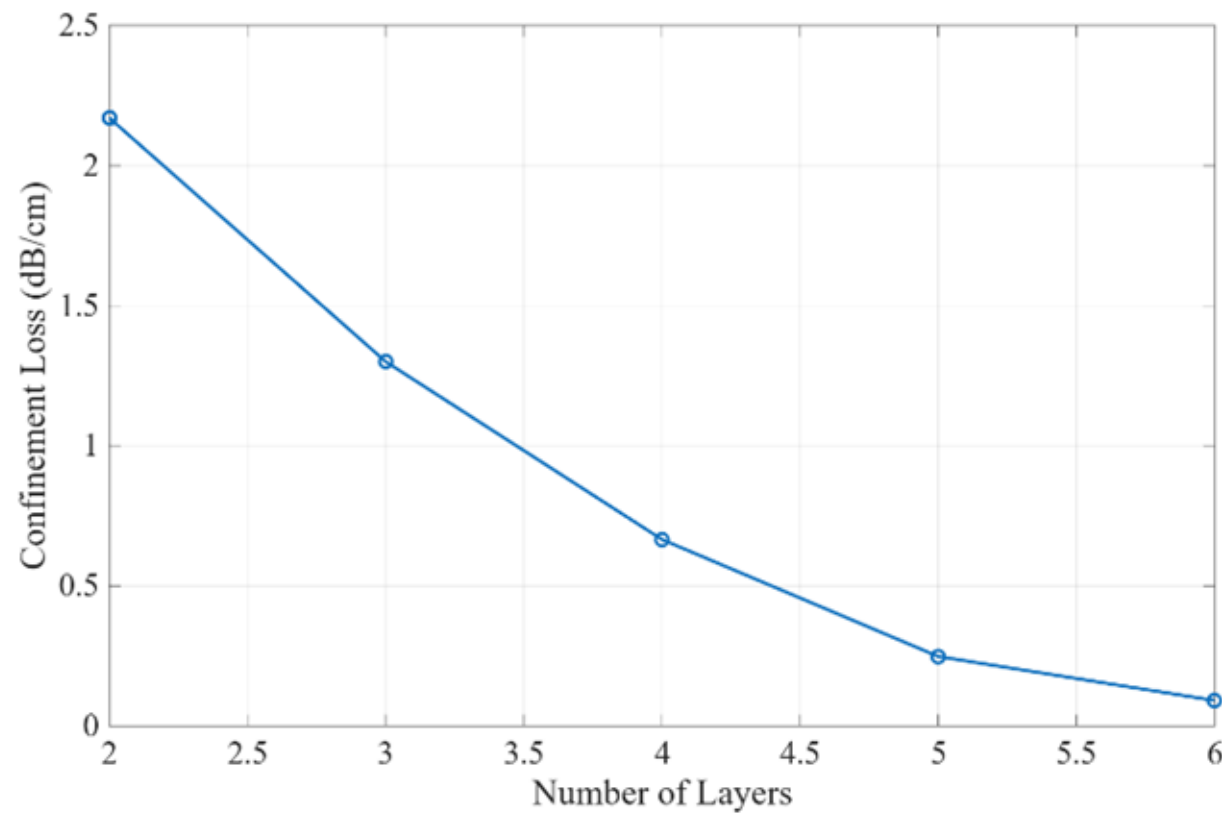

**Fig. 2.** Confinement loss versus number of cladding layers in PCF with an 8 μm lattice pitch and a core with 7 missing tracks.

A design with 7 missing tracks in the lattice was therefore chosen, with an example simulated mode profile shown in Fig. 1. To optimize this waveguide configuration, the number of cladding layers, the lattice constant, and the track refractive index were varied, to obtain the modal loss and the fundamental mode field diameter (MFD). Figure 2 shows how the confinement loss reduces as more concentric layers around the core are added. The reduction in confinement loss with increasing layer number can be attributed to the enhanced transverse confinement provided by a thicker photonic crystal cladding. As additional layers are introduced, the evanescent field is more effectively attenuated before reaching the outer boundary, thereby suppressing radiation leakage. As shown in Fig. 2, the loss reduction is particularly pronounced when the number of concentric layers increases from two to four, whereas the rate of improvement becomes smaller beyond five layers.

The effect of lattice constant on the MFD of the fundamental mode was explored, which is a key parameter for modal confinement and coupling compatibility. Figure 3 shows the variation of the fundamental-mode MFD with lattice constant for three different laser-induced refractive index changes, $\Delta n =$ -0.003, -0.004, and -0.005. For all three cases, the fundamental-mode MFD increases monotonically with increasing lattice constant, indicating that the optical field becomes progressively less confined as the lattice pitch is increased. In addition, at a fixed lattice constant, a lower refractive index contrast, $|\Delta n|$, consistently leads to a larger MFD, allowing the mode to expand further into the cladding region. These results illustrate how the lattice constant and the refractive index change affect the modal size.

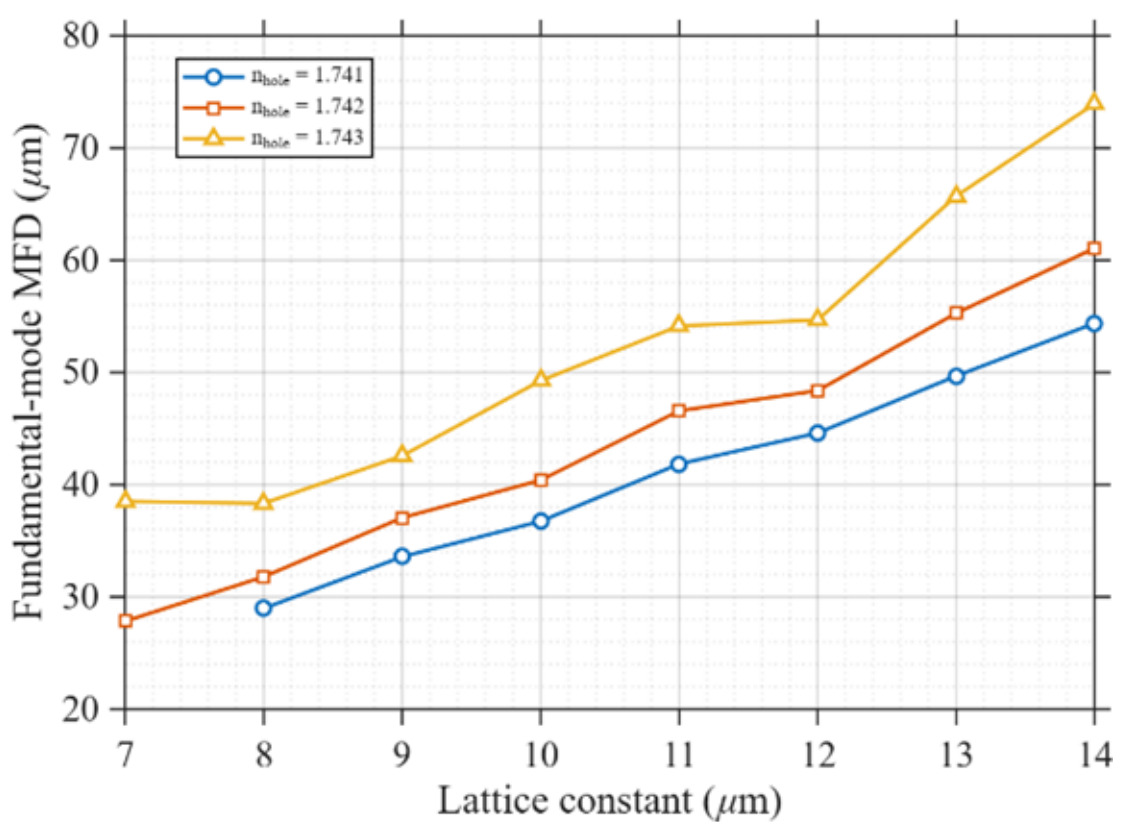

**Fig. 3.** Dependence of the fundamental-mode MFD on lattice constant for different track refractive indices. The core had 7 missing tracks.

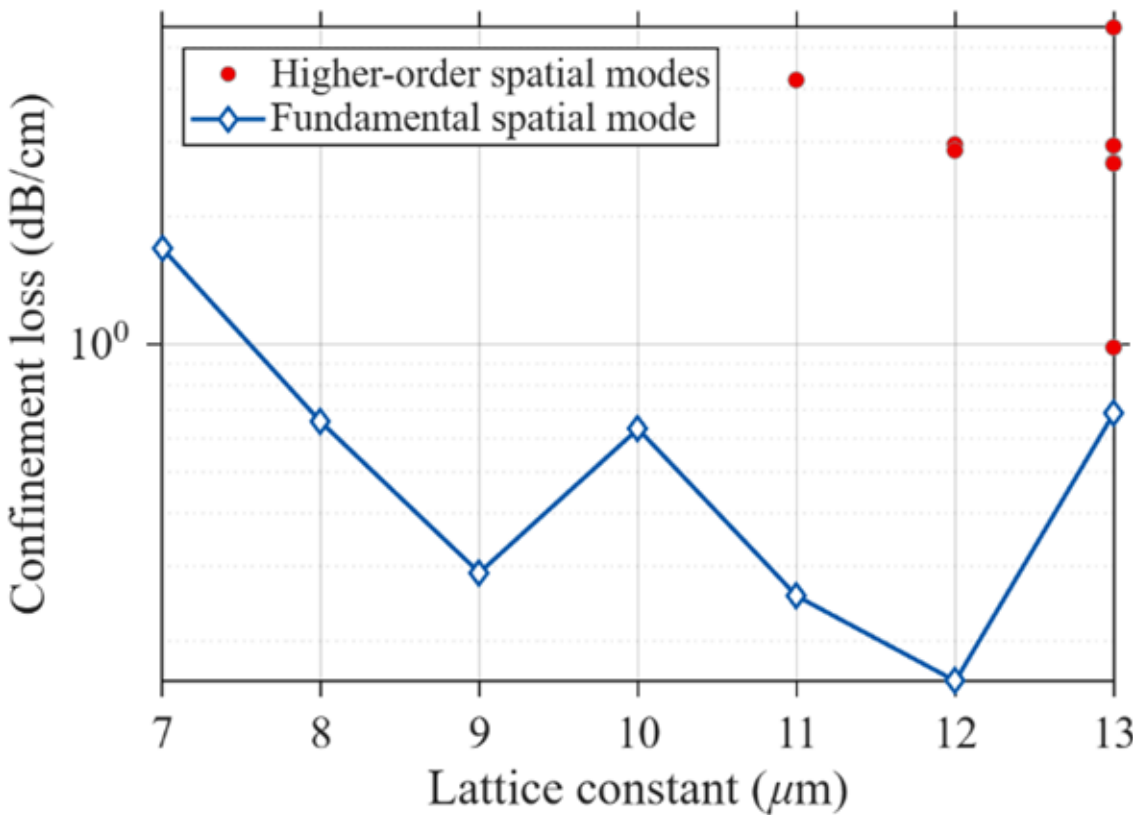

**Fig. 4.** Confinement loss of the fundamental and higher-order spatial modes versus lattice constant for a laser-induced refractive index change of -0.003. The core had 7 missing tracks.

Further analysis considers the loss behavior of the fundamental and higher-order modes, thereby providing a more complete assessment of the guiding performance and mode-selective properties of the fiber. Figure 4 compares the confinement loss of the fundamental spatial mode and higher-order spatial modes as a function of lattice constant for $\Delta n =$ -0.003 . The waveguide is single-mode for lattice constants up to 10 µm. At larger lattice constants of 11-13 µm, higher order modes become evident, but they had significantly higher loss. This loss discrimination between the fundamental and higher-order modes suggests that the structure possesses a degree of higher-order-mode suppression, which is beneficial for effective single-mode guidance.

## III. Fabrication and Optimization

### A. Fabrication system

Sapphire fibers typically exhibit a rounded hexagonal cross-section. The micromachining of a single-mode depressed cladding waveguide within a sapphire fiber has been demonstrated by measuring the cross-sectional shape, and using a spatial light modulator (SLM) to correct for the aberrations [18], [34]. However, for this new PCF design, tight control of the laser focus is required for a small spot-size with low ellipticity. The sapphire fiber shape and the radius of the top surface vary along the propagation axis [35]. This nonuniformity on a micrometer-scale, makes it challenging to precisely control the focal aberration over millimeter-scale fabrication lengths. Accordingly, an immersion oil (Cargille Labs) with a refractive index of 1.763 was used to ensure there was negligible refractive index mismatch at the interface with the fiber surface ($n_{\text{fiber}} = 1.773$ at $\lambda_0 = 515$ nm). However, conventional oil-immersion objectives are optimized for a refractive index of 1.52, leading to a mismatch with the higher-index oil. An SLM was therefore used to compensate for the aberration resulting from this mismatch.

The fabrication system used a regenerative amplified femtosecond laser system with a pulse duration of 170 fs at a second-harmonic wavelength of 515 nm. The pulse energy was controlled using a polarizer and a half-wave plate. The laser beam was expanded using a telescope and projected onto an SLM (Hamamatsu X10468), followed by a second telescope system to image the beam onto the pupil plane of a 63×, NA 1.4 oil-immersion objective (Zeiss Plan-Apochromat 440760). Sapphire samples were mounted on a three-axis nanopositioning stage (Aerotech ABL10100L and ANT95-3-V). The laser beam was linearly polarised along the optic-axis for both bulk and fiber samples.

### B. Aberration correction

The mismatch between the oil-immersion objective and the high refractive index oil used is considered as a planar interface at the objective, as illustrated in Fig. 5. The spherical aberration phase $\Phi_{\text{SA}}$ can then be mapped onto a polar coordinate system and expressed as [36]:

$$\Phi_{\text{SA}}(\rho) = -\frac{2\pi D_{nom}}{\lambda_0}\left(\sqrt{n_2^2 - (NA\rho)^2} - \sqrt{n_1^2 - (NA\rho)^2}\right) \quad (3)$$

Where $\rho$ is the normalized pupil radius, $D_{nom}$ is the nominal depth, defined as the focal depth in the absence of a refractive index mismatch. Here $n_1$ is the refractive index of conventional immersion oil (1.52), $n_2$ is the refractive index of the immersion oil used (1.763), $NA$ is the numerical aperture of the objective, and $\lambda_0$ is the wavelength of the fabrication laser.

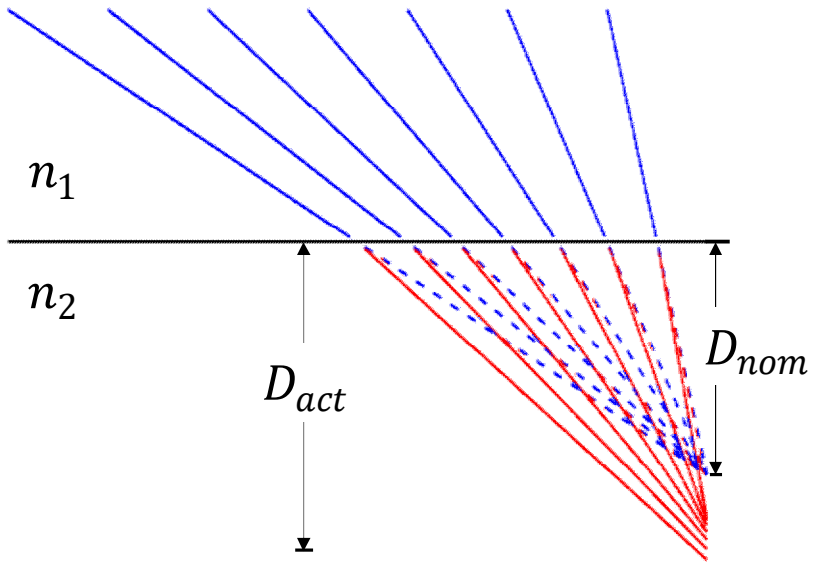

**Fig. 5.** Diagram illustrating the equivalent interface between the oil-immersion objective ($n_1$) and the high index immersion oil ($n_2$) used.

It is noted that $\Phi_{SA}(\rho)$ also includes a defocus term, which shifts the beam position from the actual focal depth, $D_{act}$, to the nominal focal depth, $D_{nom}$. To avoid complicated phase patterns, this defocus term is explicitly expressed and subtracted. The defocus term can be found using a geometric relation to be [37], [38]:

$$\Phi_{DE}(\rho) = \frac{2\pi}{\lambda_0}(D_{act} - D_{nom})\sqrt{n_2^2 - (NA\rho)^2} \quad (4)$$

An additional factor $s$ is introducted to the nominal and actual focal depths, such that $D_{\mathrm{nom}} = sD_{\mathrm{act}}$, where

$$s = \frac{n_2 - \sqrt{n_2^2 - NA^2}}{n_1 - \sqrt{n_1^2 - NA^2}} \quad (5)$$

The pre-compensated phase to be displayed on the SLM is obtained by subtracting the defocus phase $\Phi_{DE}(\rho)$ from the spherical aberration phase $\Phi_{SA}(\rho)$, yielding:

$$\hat{\Phi}_{SA}(\rho) = \frac{2\pi D_{nom}}{\lambda_0 s}\left[s\sqrt{n_1^2 - (NA\rho)^2} - \sqrt{n_2^2 - (NA\rho)^2}\right] \quad (6)$$

In this work, the sapphire fiber was fully immersed in the immersion oil with the refractive index of 1.763. This can be viewed as a scenario where the sapphire substrate is moved all the way upwards to the front lens of the objective in Fig. 5. Hence a constant aberration correction was applied across all depths during the fabrication. The objective had a working distance of 90 µm and was designed for use with a 170 µm coverslip, giving a $D_{nom} = 260$ µm. The SLM aberration correcting phase was calculated by taking the numbers using (6).

It is noted that an additional aberration arises from the refractive index mismatch between the immersion oil ($n_{\mathrm{oil}} = 1.763$) and the sapphire fiber ($n_{\mathrm{fiber}} = 1.773$). However, this contribution is significantly smaller than the aberration introduced by the mismatch between the objective design and the immersion oil. Moreover, variations in the fiber geometry along its length make this aberration difficult to quantify and compensate. This term is neglected in this work.

To test the aberration correction, laser-modified tracks were inscribed inside a sapphire bulk substrate (10 × 10 × 1.3 mm) at a depth of 100 µm below surface. The tracks were written at a repetition rate of 50 kHz and a translation speed of 4 mm/s, both with and without aberration correction. The pulse energy was varied from 10 nJ to 100 nJ. Figure 6(a, b) shows optical cross-sectional images of the resulting tracks. With aberration correction, the tracks exhibit an improved aspect ratio, particularly at higher pulse energies, as shown in Fig. 6(c). The laser modification threshold is lower for the aberration-corrected tracks, due to the tighter focal confinement.

The reduced track dimensions are advantageous for transferring the design to an optical fiber platform, where compact geometry is required. Smaller tracks enable more compact devices with closer feature spacing and reduced likelihood of crack formation. Figure 6(d, e) shows representative cross-sectional images of the fabricated photonic crystal fibers with and without aberration correction, using the same pulse energy. Without aberration correction, the elongated tracks are more prone to crack formation and exhibit greater anisotropy between the horizontal and vertical directions, which could lead to larger birefringence.

**(a)** Tracks fabricated without Aberration correction

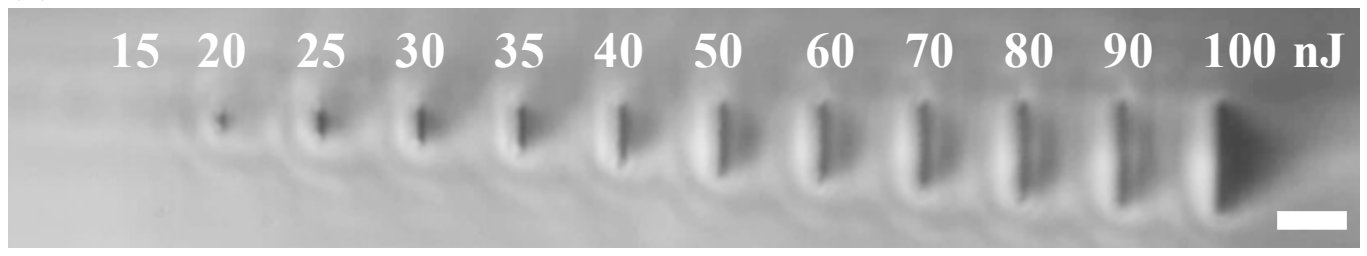

**(b)** Tracks fabricated with Aberration correction

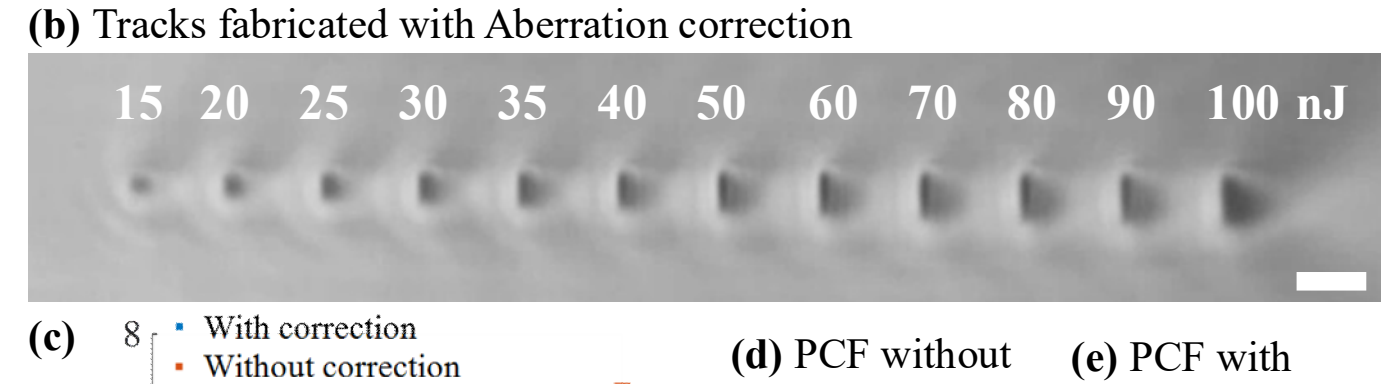

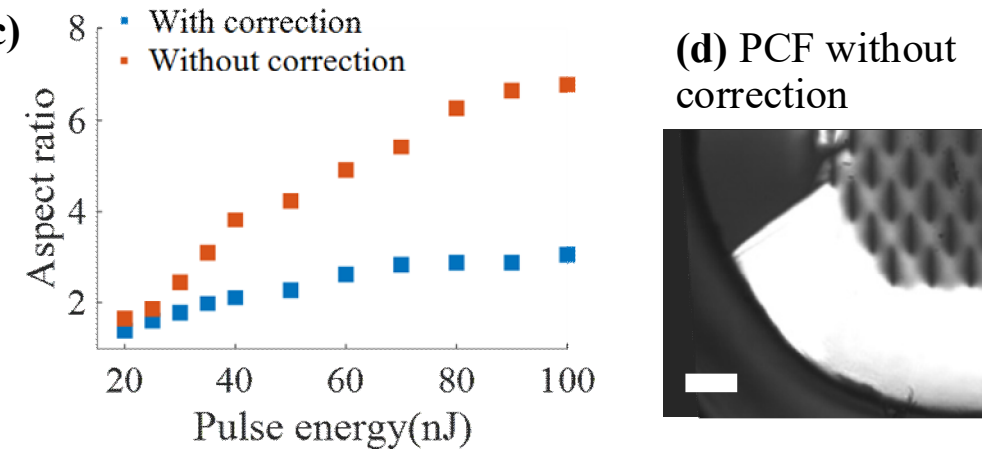

**(d)** PCF without correction

**(e)** PCF with correction

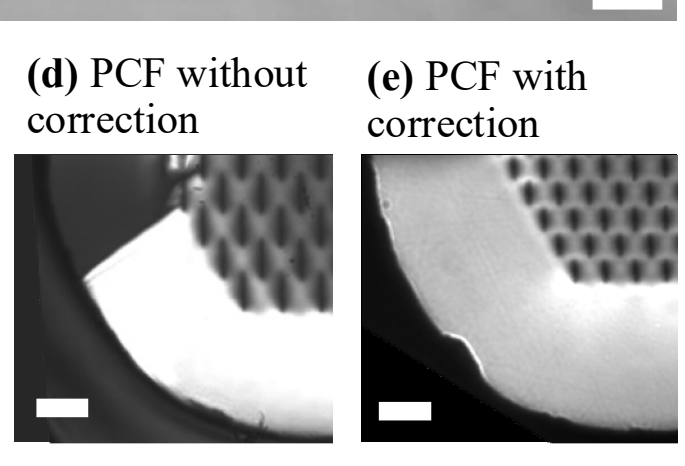

**Fig. 6.** (a, b) Cross-sectional images of laser-modified tracks fabricated with pulse energies ranging from 15 to 100 nJ, (a) without aberration correction and (b) with aberration correction. (c) Aspect ratio of the tracks, calculated from the measured height and width as a function of pulse energy. (d, e) Cross-sectional images of a 150-µm diameter sapphire photonic crystal fiber, fabricated (d) without and (e) with aberration correction**.** The white scale bars represent 15 µm.

### *C. Optimization of the photonic crystal waveguide and fabrication conditions*

With the guidance from the simulation results in Section II, experimental optimization of the sapphire photonic crystal waveguide was carried out on bulk sapphire substrates. The influence of laser fabrication conditions, particularly the pulse energy, as well as waveguide geometry, were investigated. Waveguiding performance, including the MFD, mode field mismatch loss, and propagation loss, were evaluated. Attention was also given to device compactness, fabrication time, and the absence of cracks, as these are critical for scalable sapphire photonic crystal fiber sensor integration and stable long-term operation.

The fabrication speed was fixed at 4 mm/s. This is faster than our previous work [19], and further improvement may be possible. The repetition rate was set to 50 kHz. Waveguides were fabricated at 100 µm below the sample surface. The oil-immersion objective with refractive index matching oil ($n_{\mathrm{oil}} =$ 1.763) was used, and aberration correction was applied to all samples. After fabrication, the samples were polished to optical quality to remove the edge chamfer and expose the waveguides.

For mode field characterization, 1550 nm light was launched from a single-mode fiber (Corning SMF-28e+ or equivalent), mounted on a six-axis alignment stage, and butt-coupled to the sapphire waveguide. The near-field mode profile was measured using an imaging system and an NIR camera.

Photonic crystal waveguides with four lattice layers and a total of 84 modification tracks were fabricated in a 1-cm-long sapphire substrate. Both pulse energy and lattice constant were varied, with pulse energies ranging from 30 to 80 nJ and lattice constants ranging from 4 to 12 µm. The fabricated waveguides were annealed at 1200 °C for 1 hour by heating to 1200 °C at a rate of 5 °C/min, followed by natural cooling, in order to relieve residual stress and improve high-temperature stability [39]. Figure 7(a–c) shows optical cross-sectional images of the waveguides with three different lattice constants. The corresponding mode profiles are shown below each waveguide.

The MFD was extracted from the profiles and plotted in Fig. 7(d). The MFD of a standard single-mode fiber (10.4 µm) is also indicated by the black dashed line. Optimized mode field matching could be achieved using a photonic crystal waveguide with a lattice constant of 4 µm, fabricated with pulse energies of 30 nJ or 45 nJ. The corresponding mode field mismatch loss was calculated to be less than 0.1 dB. However, cracks were observed along the lattice due to the dense track spacing, which is undesirable as it can lead to mechanical instability.

As shown in Fig. 7(d), the mode field size increases with increasing lattice constant, leading to a higher mode field mismatch with a silica fiber. The mode field mismatch losses for waveguides with a 6 µm lattice constant were calculated to be between 0.6 and 0.7 dB, for pulse energies between 30 and 80 nJ. The corresponding losses for an 8 µm lattice constant were between 1.5 and 2.7 dB. For the same lattice constant, the MFD decreases with increasing pulse energy. This is attributed to improved modal confinement resulting from the higher refractive index contrast introduced at higher pulse energies. The total insertion loss (TIL) was measured by comparing the transmitted optical power from the waveguide with that directly from a reference optical fiber. The results are shown in Fig. 7(e). The TIL can be expressed as

$$\mathrm{TIL(dB)} = \alpha(\mathrm{dB/cm})z(\mathrm{cm}) + L_{coupling}(\mathrm{dB}) \quad (7)$$

where $\alpha$ is the propagation loss in dB/cm and $z$ is the waveguide length. The propagation loss includes both mode confinement loss and scattering loss due to inhomogeneities inside the waveguide. The coupling loss is given by

$$L_{coupling} = L_{mismatch} + L_{system} + L_{Fresnel} \quad (8)$$

where the terms correspond to mode-field mismatch, system losses due to angular or lateral misalignment and other imperfections, and Fresnel reflections at the fiber–air and air–sapphire interfaces. Regardless of lattice constant, increasing the pulse energy from 30 to 50 nJ reduces the total insertion loss. This is likely due to reduced propagation loss, since higher pulse energy increases the refractive index contrast and therefore improves mode confinement. For a lattice constant of 8 µm, the improvement becomes less pronounced above 50 nJ. A crack-free waveguide was found for lattice constants of 6-8 µm using a pulse energy of around 45 nJ. The design was confirmed to be single-mode by varying the input coupling conditions to excite higher-order modes.

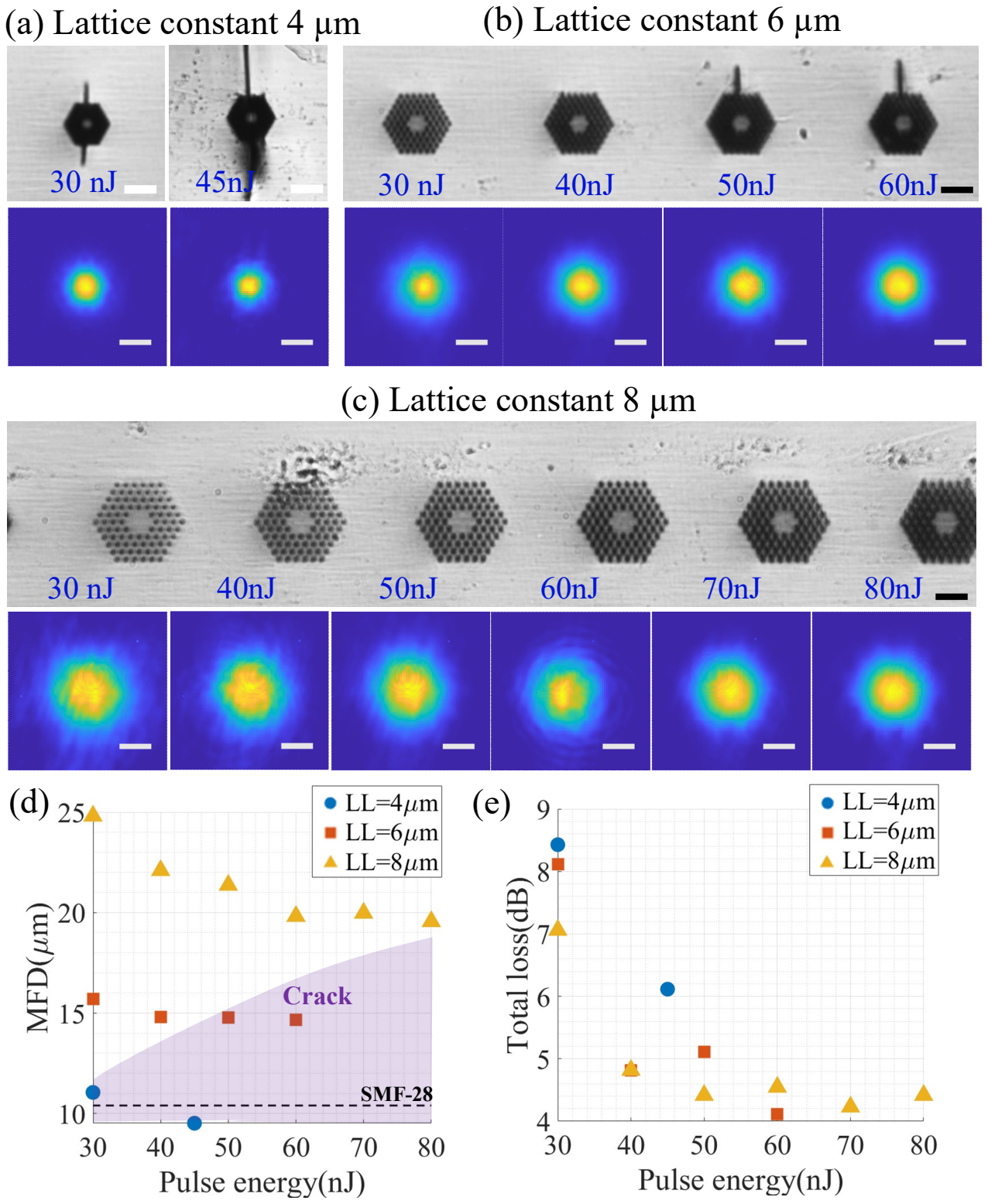

**Fig. 7.** (a-c) Microscope cross-sectional images of photonic crystal waveguides with a lattice constant of (a) 4 µm, (b) 6 µm, and (c) 8 µm after annealing at 1200 °C, fabricated with pulse energies ranging from 30 to 80 nJ, and the measured 1550 nm mode profiles of the corresponding waveguides. The scale bars on the microscope images represent 20 µm. The scale bars on the mode profiles represent 10 µm. (d) MFD of the waveguides in (a-c). The markers in the shaded region exhibit cracks. (e) Measured total insertion loss of the waveguides as a function of pulse energy for different lattice constants.

The impact of the number of lattice layers was also investigated. Photonic crystal waveguides with 2 to 5 layers were fabricated using a pulse energy of 45 nJ and a lattice constant of 8 µm. Cross-sectional images in Fig. 8(a) show that the waveguides remain crack-free up to four layers. Cracks begin to appear when five layers are used, even at larger constant sizes, due to the increased structural complexity. The measured mode profiles in Fig. 8(b) show minimal variation with layer number, and consequently the mode field mismatch loss [Fig. 8(c)] remains largely unchanged. In contrast, the total insertion loss [Fig. 8(c)] decreases with increasing layer number, likely due to improved confinement and reduced radiative loss. This indicates a trade-off between fabrication time and optical performance. The fabrication time is approximately 2 minutes per centimeter of waveguide for the two-layer design, increasing to about 7 minutes per centimeter for the five-layer waveguide [Fig. 8(d)].

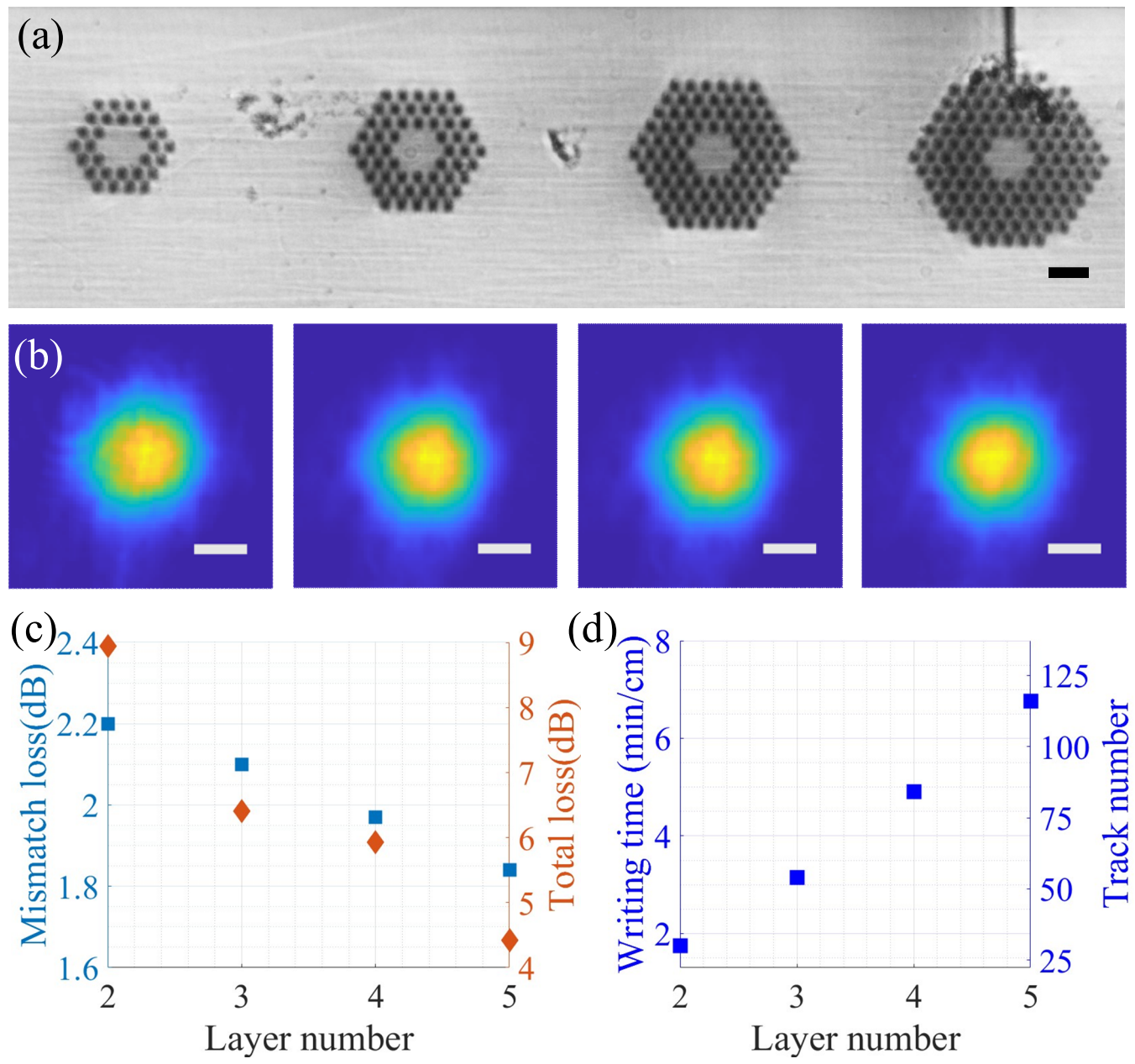

**Fig. 8.** (a) Microscope images of photonic crystal waveguides after annealing at 1200ºC with varying outer layer numbers between 2 to 5, the lattice constant is 8 µm; the scale bar represents 20 µm; (b) the corresponding measured 1550-nm mode fields of the photonic crystal waveguides; the scale bars represent 10 µm; (c) the mode field mismatch loss (blue square) and total loss (orange diamond) dependence on the layer number, and (d) the writing time and track number dependence on the number of layers.

## IV. Sapphire Fiber Sensor Fabrication

### A. Photonic Crystal Waveguide Inscription

Single crystal sapphire fiber 150 µm in diameter, was fabricated using a custom laser heated pedestal growth system [40], [41]. The sapphire fiber was cut into lengths of 4 cm and 7 cm. Within each fiber a photonic crystal waveguide was inscribed with a lattice constant of 8 µm, a core formed of 7 missing tracks, and 4 concentric layers of tracks around the core. A pulse energy of 45 nJ, a writing speed of 4 mm/s, and a repetition rate of 50 kHz were used. The fabrication conditions and waveguide geometry were selected based on the simulation and experimental results on the bulk sapphire. Adjustments were made heuristically to achieve optimized fiber sensor performance. After fabrication, the sapphire PCFs were polished on one side to optical quality to expose the waveguide for fusion splicing. The sensor was annealed at 1000°C for an hour after fabrication and before fusion splicing.

Figure 9 shows the cross-section and mode field profile of resulting photonic crystal fiber. The cross-sectional image is shown in Fig. 9(a). The fiber remains completely crack-free, even after polishing, which is attributed to the precise aberration correction. The mode field is shown in Fig. 9(b). The final design selected for the sapphire PCF therefore employs four layers and a lattice constant of 8 µm. This design was chosen over a lattice constant of 6 µm, as the fiber is more vulnerable to cracks than the bulk, due to its reduced volume. The mode field mismatch loss with standard optical fiber is approximately 2 dB, though this could be reduced by using a mode converter. The design consists of 84 modification tracks, written at a laser-writing speed of 4 mm/s. The fabrication time to make the full structure is approximately 5 minutes per centimeter of waveguide. This is a 6-fold improvement on the fabrication time of our depressed cladding waveguides [19], allowing a significant improvement in the cost of manufacture.

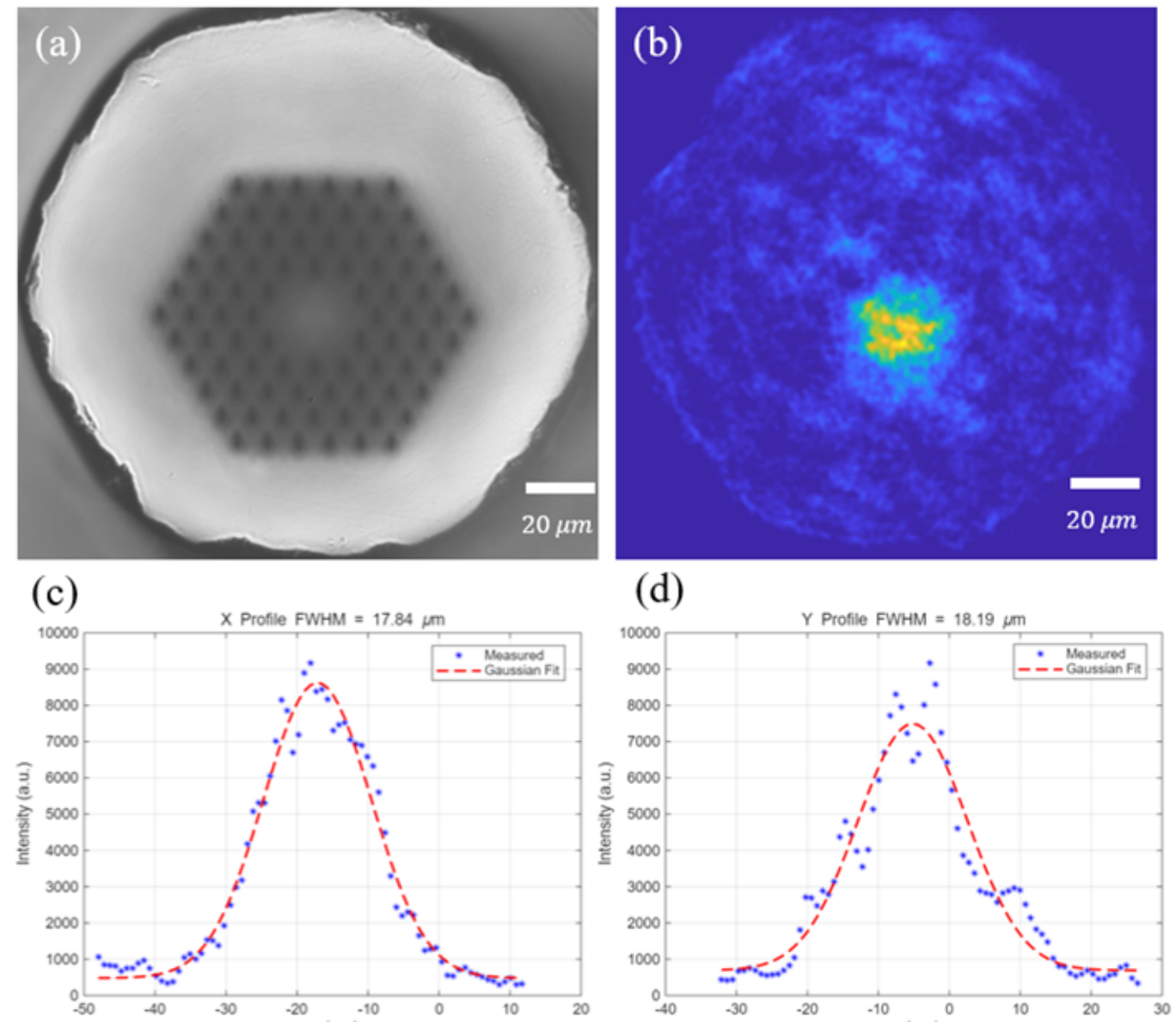

**Fig. 9.** (a) Cross-sectional image of the photonic crystal fiber. (b) The experimentally measured 1550-nm mode field for a 7-cm long sapphire photonic crystal fiber after annealing. (c) Mode field intensity profile and Gaussian fit along the horizontal direction in the fiber core region. (d) Mode field intensity profile and Gaussian fit along the vertical direction in the fiber core region.

Figure 9(c-d), shows the intensity distribution within the fiber core and Gaussian profiles for comparison, with reasonable agreement. This indicates that the guided mode in the core region is close to a single-mode Gaussian distribution, although minor background and peripheral features can still be observed. The speckle pattern could be a result of non-homogeneity in the laser-written tracks in the sapphire fiber.

The photonic crystal fiber attenuation was measured using the cut-back method, in which the output power is recorded at different fiber lengths under identical input conditions. Figure 10 shows the power in dBµW against length and a linear fit. There is considerable uncertainty using the cut-back method with waveguides in crystals as the coupling losses vary significantly with the quality of the polish, hence there is scatter in the data. However, the linear regression yields an estimate of the propagation loss of 0.7 dB/cm. This is around half the value of our previous design [19]. The coupling loss could be improved with mode-matching between the silica fiber and the sapphire fiber.

### B. Fiber Bragg Grating

Fiber Bragg gratings (FBGs) were inscribed within the photonic crystal waveguides. The gratings were second-order, with a period of 887.76 nm. The inscription was performed using a burst-pulse method with a pulse energy of 60 nJ, a writing speed of 1 mm/s, and a repetition rate of 50 kHz, corresponding to 44 pulses per grating period. The duty cycle was set to 10%. The gratings had a length of 5 mm and were positioned towards one end of the fiber. A cross-sectional image of the sapphire FBG is shown in Fig. 11, where the grating is shown at the center of the waveguide. However, for the fiber sensors, the grating was terminated before reaching the fiber end to mitigate interference fringes arising from a Fabry-Perot cavity.

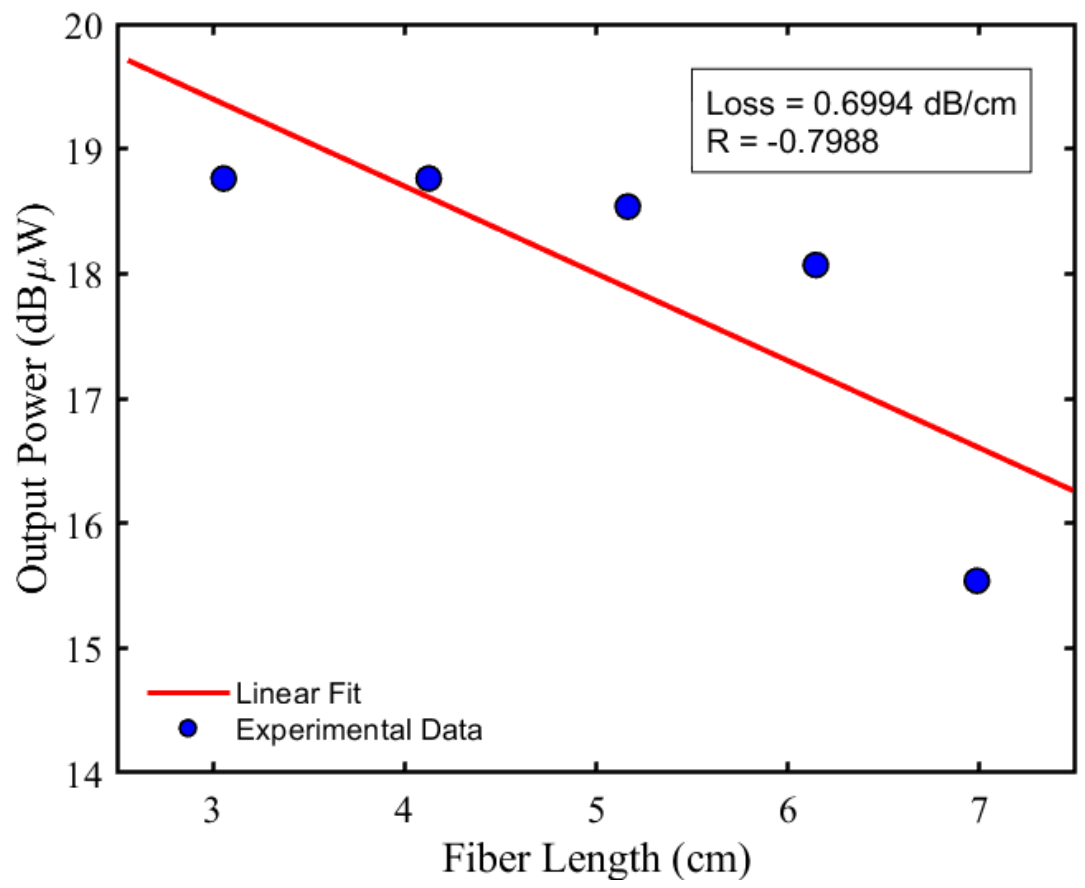

**Fig. 10.** Loss measurement and fitting results using the cut-back method.

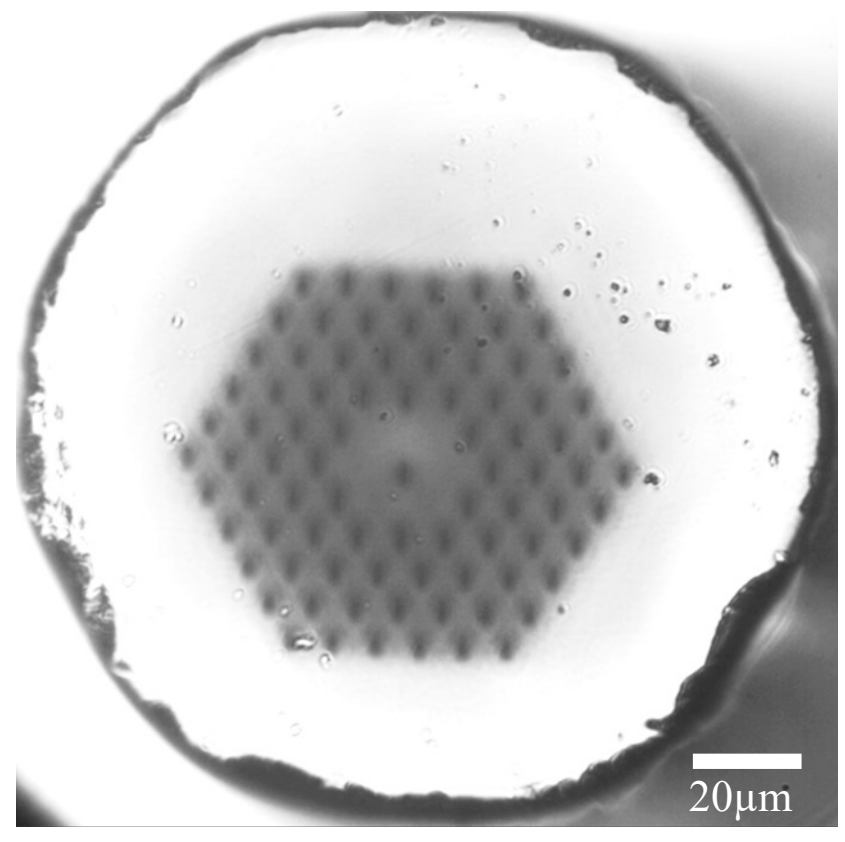

**Fig. 11.** The cross-sectional image of a 7-cm sapphire FBG, showing the grating in the middle.

### *C. Fusion splicing*

To realize a practically deployable sapphire photonic crystal FBG temperature sensor, the sapphire fiber was fusion spliced to a standard single-mode silica fiber (Corning SMF-28e+, or equivalent). Since the sapphire fiber has a rounded hexagonal outer profile and a diameter of 150 µm, which is slightly larger than the 125 µm cladding of SMF-28e+, the manual alignment mode of a Sumitomo Type-72C splicer was used. During alignment, the single-mode silica fiber was connected to a commercial FBG interrogator (Halliburton/SmartFibres SmartScan SBI). The interrogator's real-time reflection spectrum was monitored to locate the position yielding the strongest Bragg peak, thereby achieving near core-to-core alignment before initiating the splice. During splicing, the arc time and prefuse time were set to 3 s, with the arc and prefuse power set to the same values used for splicing silica-core fibers.

Despite the mismatch in diameter and cross-sectional geometry, the fusion joint passed the splicer's automatic proof (tensile) test. The fusion-spliced joint exhibited good mechanical robustness under bending and handling. Photographs of the fiber after splicing are shown in Fig. 12. The FBG reflection spectrum of the spliced assembly the Bragg peak power was reduced relative to the pre-splice level. A slow-motion review of the splicing video indicates that the splicer momentarily separates the fiber ends prior to arc discharge and then retracts them, introducing a slight positional shift that degrades core alignment at the moment of fusion and causes a reduction in the peak power.

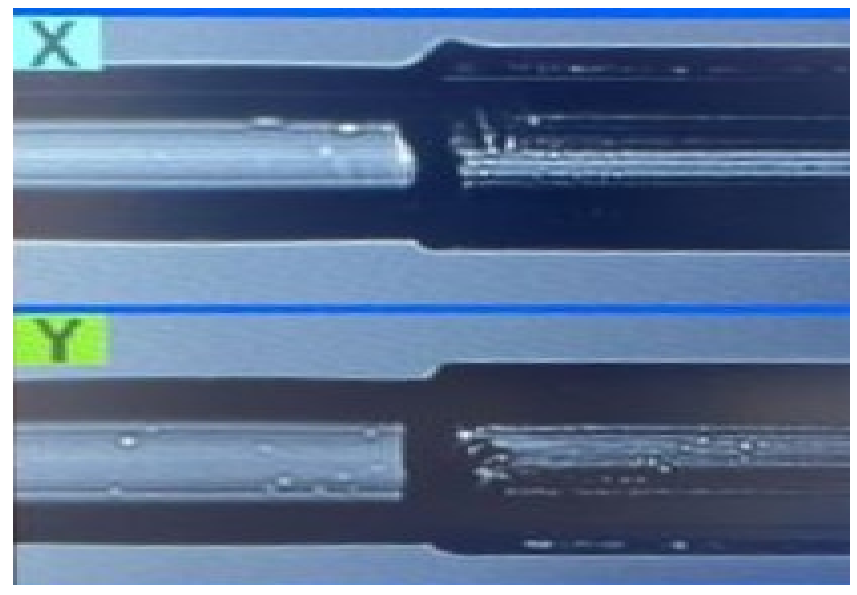

**Fig. 12.** Splicer microscope images of the joint after fusion splicing. The 125-µm-diameter SMF-28e+ fiber is on the left and the 150-µm-diameter sapphire FBG sensor is on the right. The photonic crystal waveguide can be identified in the image.

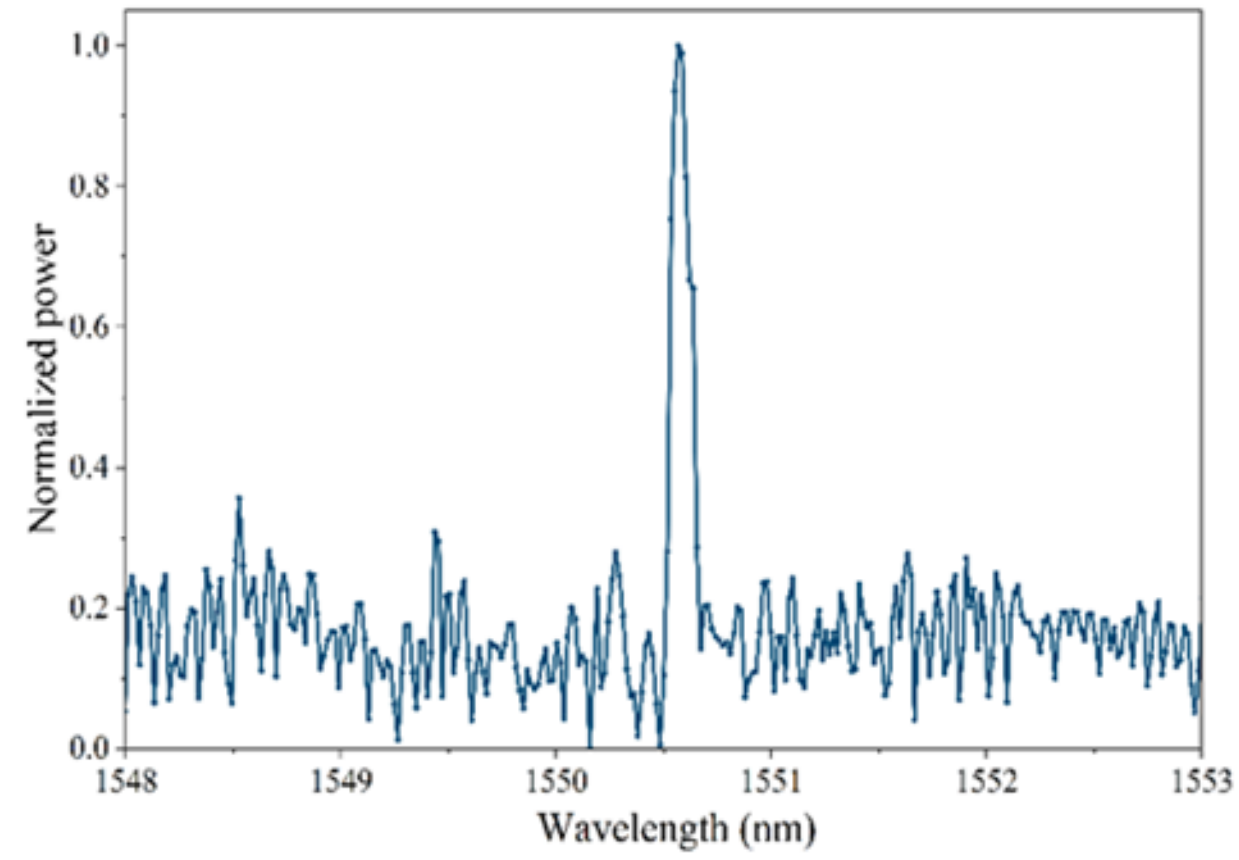

**Fig. 13.** The reflectivity spectrum of the 7 cm sapphire FBG sample after fusion splicing.

The post-splice FBG spectrum of a 7-cm long sapphire temperature sensor is shown in Fig. 13. The Bragg wavelength was measured to be 1550.6 nm, with a full-width at half-maximum (FWHM) of approximately 0.12 nm. The narrow and well-defined reflection peak suggests predominantly single-mode-like spectral behavior, although higher-order mode contributions cannot be fully excluded.

## V. Temperature Sensing Performance

The 7 cm sensor was placed in a box furnace and heated to 700 ℃ in 100 ℃ increments, while an interrogator recorded the FBG reflection spectrum in real-time. At each temperature setpoint, multiple spectra were acquired and averaged to suppress jitter. The normalized results are shown in Fig. 14, revealing a nonlinear dependence of the Bragg wavelength on temperature.

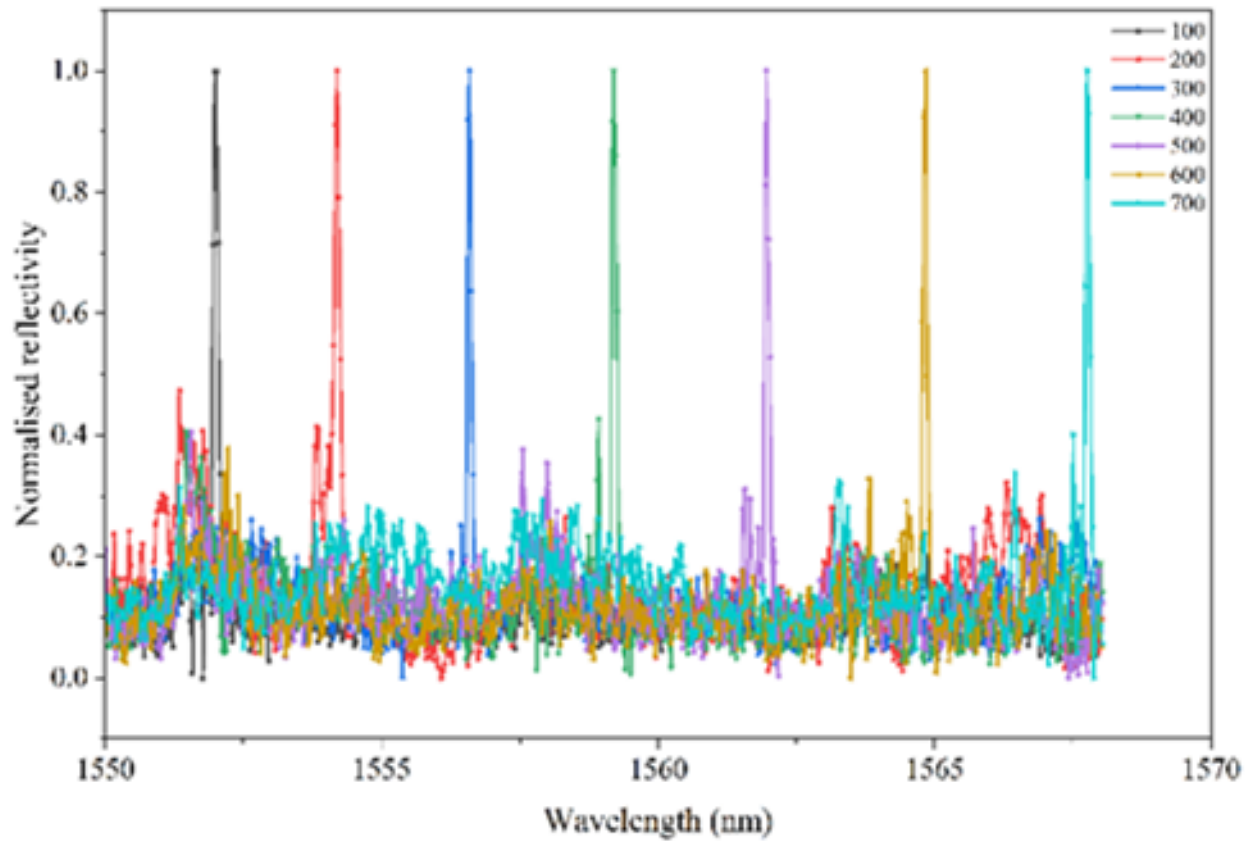

**Fig. 14.** Spectra of the 7 cm sapphire FBG sensor at different temperatures between 100 ℃ to 700 ℃.

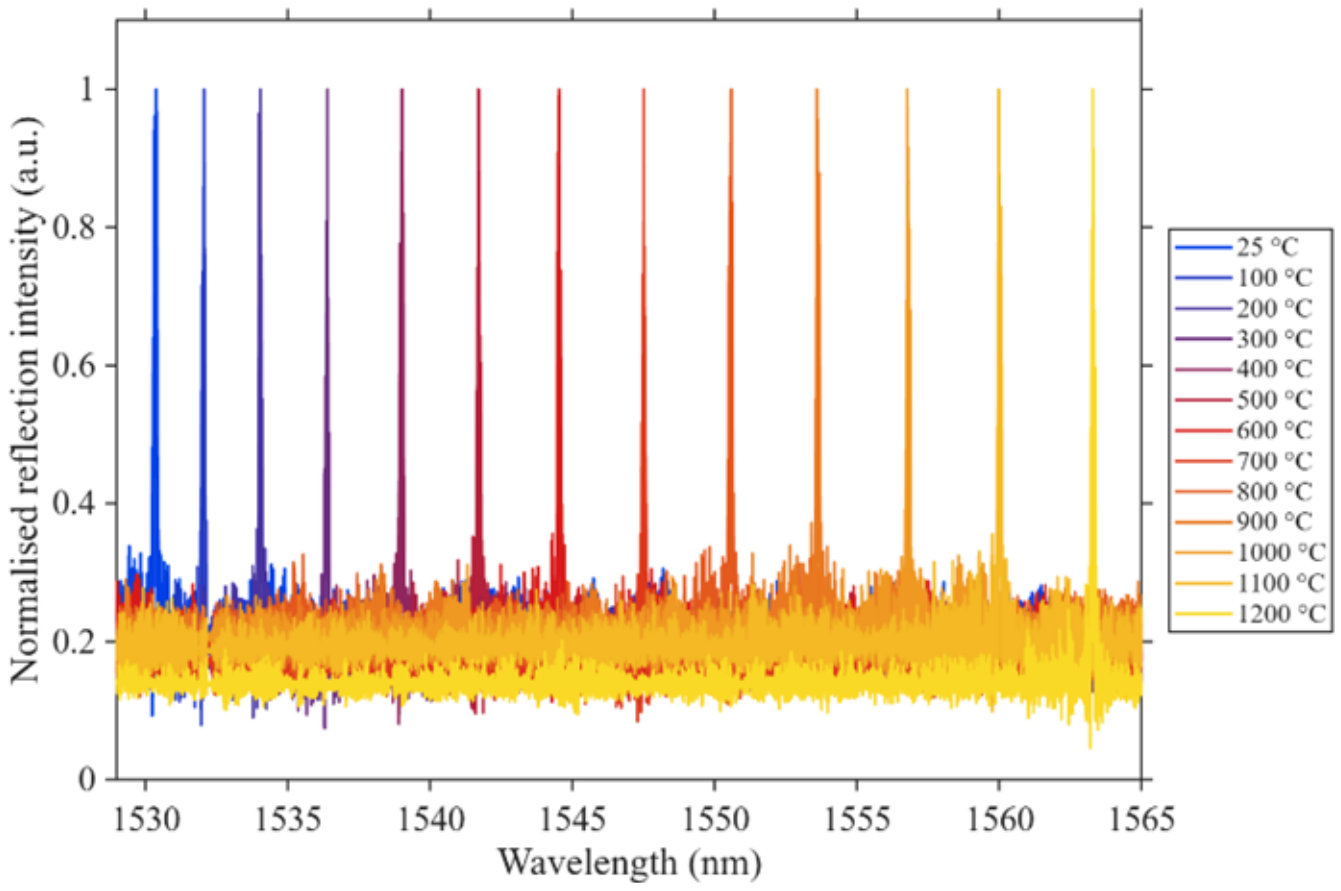

**Fig. 15.** Spectra of the 4 cm sapphire FBG sensor at different temperatures.

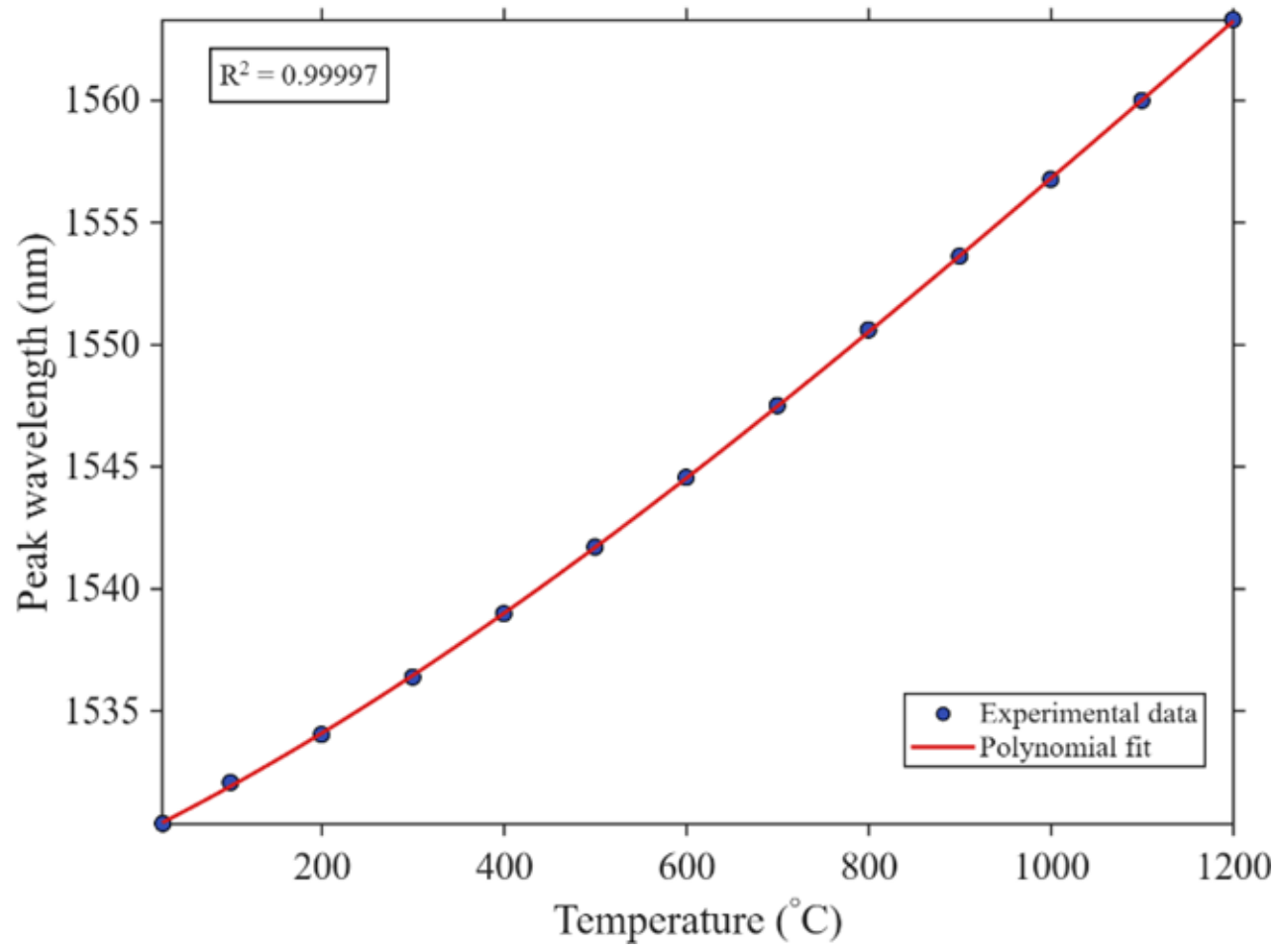

**Fig. 16.** Characterization curve of the 4 cm sapphire FBG sensor at different temperatures.

Due to the limited yield of the 7 cm samples, multiple 4 cm long samples were also fabricated for high-temperature testing in order to obtain sufficient experimental data. Figure 15 shows the normalized reflection spectra of the 4 cm fiber sensor over the temperature range from 25 °C to 1200 °C. As the temperature increases, the reflection peak exhibits a clear and continuous red shift, indicating a stable and pronounced wavelength response to temperature.

Figure 16 presents the relationship between the extracted peak wavelength and temperature. The peak wavelength increases monotonically with temperature and exhibits a slightly nonlinear dependence. A third-order polynomial fit agrees very well with the experimental data, yielding an excellent coefficient of determination of $R^2 = 0.9999$. The temperature sensitivity was 19.0 pm/°C at 25 °C, rising to 32.3 pm/°C at 1200 °C.

## VI. Conclusions

This work has demonstrated a sapphire photonic crystal fiber (PCF) temperature sensor. An index-guiding photonic crystal waveguide was written into coreless 150 µm single-crystal sapphire fiber using femtosecond laser direct writing. Within this waveguide a 5 mm second-order Bragg grating was written. A high-index immersion oil matched to the sapphire was used to mitigate the nonuniform fiber surface. An SLM was used to compensate for aberration introduced by the mismatch between this oil and a standard immersion-oil objective. The sapphire fiber was fusion spliced to a lead-in silica fiber, enabling compatibility with standard interrogation systems. Devices up to 7 cm were long were fabricated, greatly expanding the range of applications for ultra-high temperature sensing. The propagation loss was estimated to be 0.7 dB/cm, giving significant potential for extending the length further. The Bragg grating bandwidth was approximately 0.12 nm. The devices were tested over a temperature range between 25 to 1200 °C, showing a corresponding temperature sensitivity of 19.0–32.3 pm/°C across this range. The simplified photonic crystal geometry allows the full waveguide to be written at speeds of 5 min/cm. This is approximately a 6-fold improvement in the fabrication time, significantly reducing the cost of manufacture. It also reduces the propensity to cracking, thereby improving reliability and lifetime. These results provide a pathway towards high-precision, remote single-mode sapphire fiber sensors for many extreme-environment applications

## Acknowledgment

The authors gratefully acknowledge the support and advice of their partners Rolls-Royce plc, Cranfield University, UK Atomic Energy Authority and MDA Space and Robotics. The authors thank Prof Felix Hofmann for the use of the furnace. They thank Professor Dominic O'Brien, Dr Gabriel Araneda Machuca for the use of equipment.

# Sapphire Photonic Crystal Fiber Sensor (Supplementary Material)

Mohan Wang, Tongyu Liu, Zipei Song, Richard Reeves, Frank P. Payne, Igor N. Dyson, Kaihui Zhang, Tao Wang, Jian Zhang, Zhitai Jia, Patrick S. Salter, Martin J. Booth, *Fellow, Optica*, and Julian A. J. Fells, *Member, IEEE*

## Introduction

The main paper describes a sapphire photonic crystal fiber created using femtosecond laser direct writing. It was found that a lattice structure incorporating a core of 7 missing tracks was the optimum configuration. However, in reaching that conclusion, work was also performed to assess a structure with a core consisting of a single missing track. This Supplementary Material provides a summary of this work.

## Simulation

The performance of a photonic crystal fiber waveguide design in which only the single central track is omitted was simulated using COMSOL Figure S1 shows the simulated mode field at 1550 nm. The photonic crystal waveguide has a lattice constant of 8 µm and there are 5 concentric layers of tracks around the core. The inscribed tracks have a refractive index change of -0.003, which was estimated from the post-annealed mode profile of the actual waveguide. The mode field diameter is much smaller than that of the optimised design demonstrated in the main paper, due to the reduced guiding area size. However, there is considerable leakage into and beyond the lattice cladding structure.

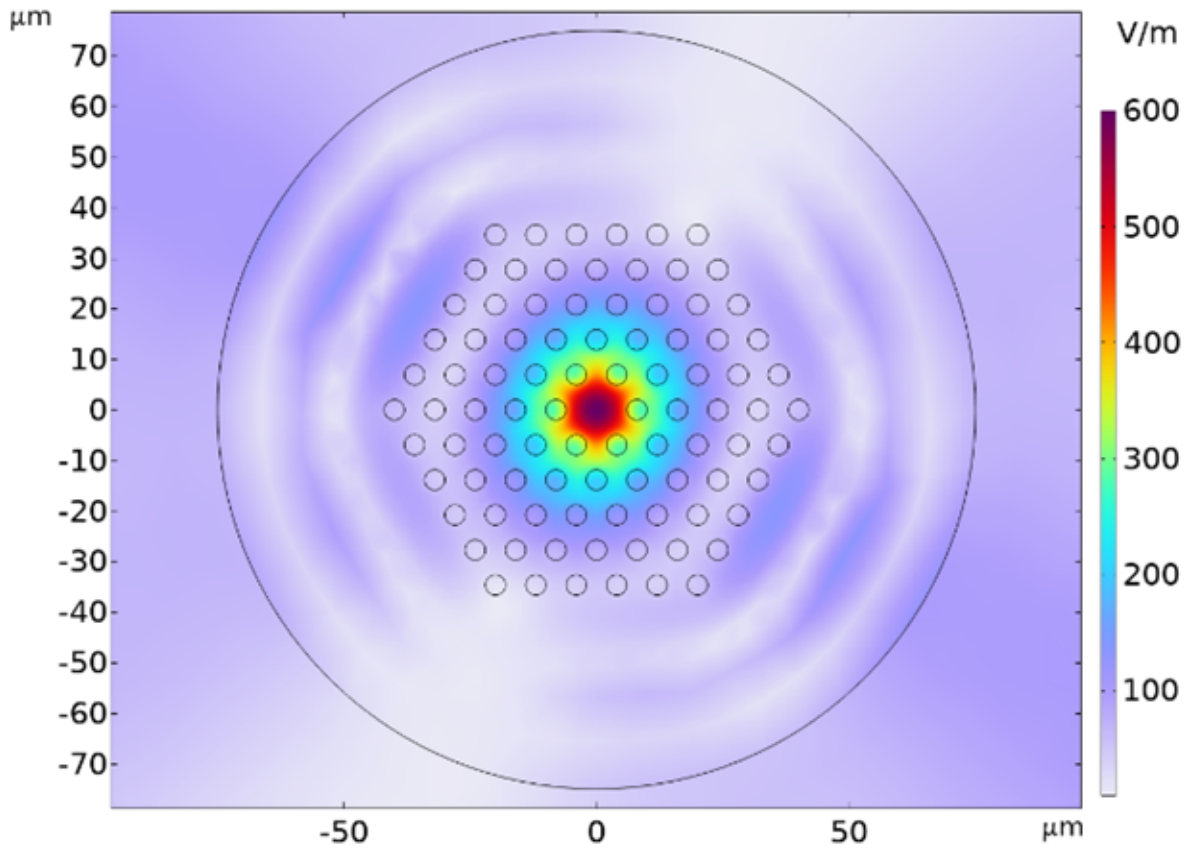

**Fig. S1.** Simulated mode profile of a PCF waveguide with only the central track missing.

## Fabrication in bulk sapphire

The photonic crystal waveguide design was then fabricated on a 1-cm long bulk sapphire substrate, using a repetition rate of 50 kHz, a pulse energy of 30 nJ, and a writing speed of 4 mm/s. The rest of the fabrication conditions were the same as those described in the main paper. Figure S2 shows microscope cross-sectional images and the 1550-nm mode fields for lattice constants between 4 and 6 µm. For a lattice constant of 4 µm, no efficient guiding could be observed, as the core size was too small to support low loss mode propagation. Fundamental modes could be observed for lattice constants of 6 and 8 µm; however, a considerable portion of the light leaked into the nearby lattices, leading to increased loss.

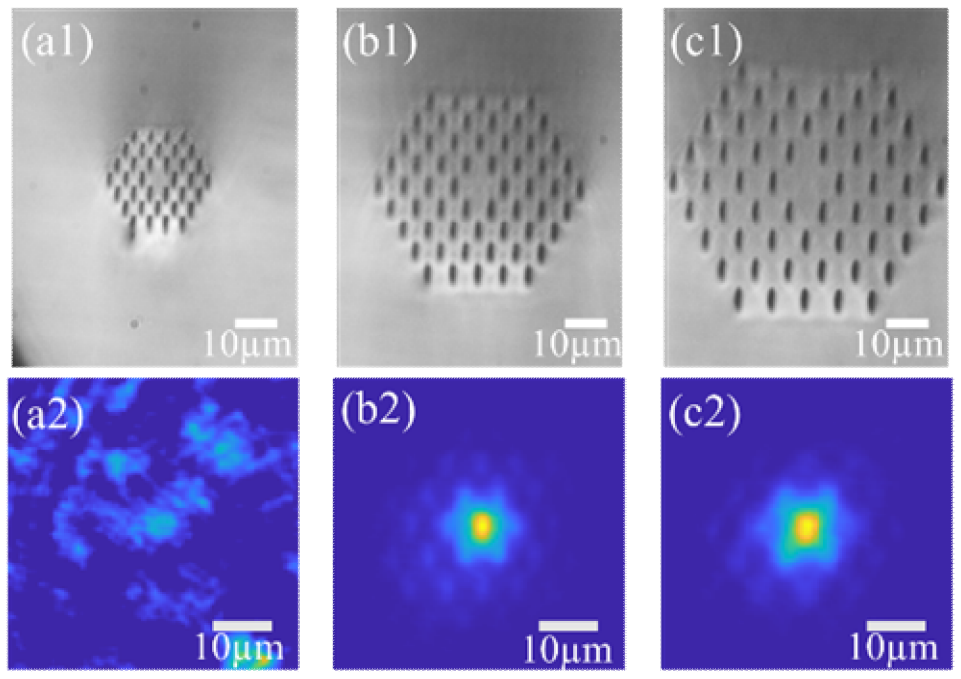

**Fig. S2.** Cross-sectional images and mode fields of photonic crystal waveguides fabricated on bulk sapphire with lattice constants of (a) 4 μm, (b) 6 μm, and (c) 8 μm.

## Fabrication in sapphire fiber

The photonic crystal waveguide was fabricated on a 150-μm-diameter sapphire fiber using a pulse energy of 45 nJ. Figure S3 shows the cross-sectional microscope image and the mode field profile of the photonic crystal waveguide. The speckle pattern observed in the fiber mode field, compared with the bulk waveguide mode field, was likely a result from non-homogeneity in the laser-written tracks. Cracks could be observed outside the cladding, as well as within the waveguide cladding, including the innermost layer, due to the dense spacing of the tracks.

Attempts to inscribe a Bragg grating within this design were unsuccessful. The additional grating in the center of the waveguide would further decrease the effective refractive index of the core, reducing the refractive index contrast between the core and the weakly guiding cladding, and thus degrading the overall guiding performance. However, with the smaller mode field, this design could still potentially function as a single-mode waveguide for light delivery.

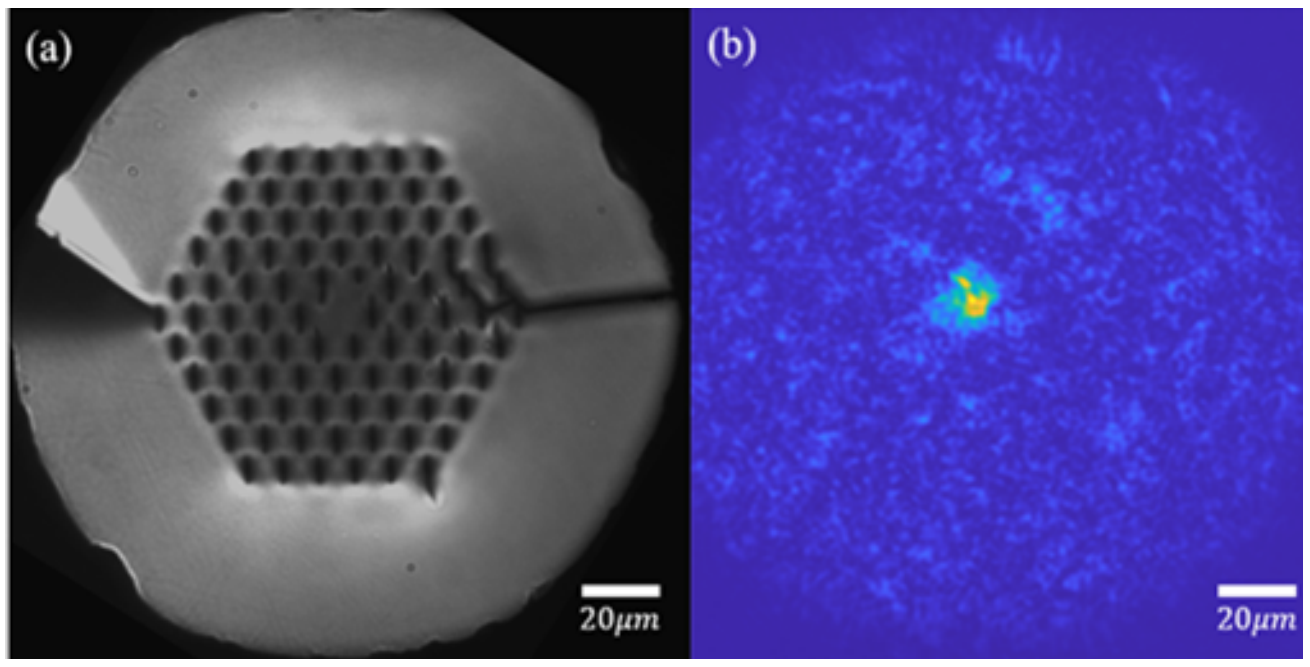

**Fig. S3.** (a) Microscope image of polished end face and (b) experimentally measured 1550-nm mode field for a 4-cm sapphire photonic crystal fiber.